\documentclass[aps,a4paper,10pt,twocolumn,nofootinbib]{revtex4} 
\usepackage[T1]{fontenc}
\usepackage{mathptmx}
\usepackage{datetime}
\usepackage{amsmath}
\usepackage{amsfonts}
\usepackage{amsfonts}
\usepackage{mathrsfs}
\usepackage[mathscr]{euscript}
\usepackage[dvips]{graphicx}
\usepackage{fancyhdr}
\usepackage{colordvi}
\usepackage{hyperref} 
\usepackage{epsfig}
\usepackage{color}
\usepackage{bm}

\def\overstrike#1#2{{\setbox0\hbox{$#2$}\hbox to \wd0{\hss
    $#1$\hss}\kern-\wd0\box0}}

\renewcommand{\Vec}{\bm}

\newdateformat{yymmdddate}{\THEYEAR/\twodigit{\THEMONTH}/\twodigit{\THEDAY}}

\newcommand{\XDOI}[1]{\href{http://dx.doi.org/#1}{doi:#1}}
\newcommand{\XARXIV}[1]{\href{http://arxiv.org/abs/#1}{arXiv:#1}}

\newcommand{\pTime}{t}
\newcommand{\pSpace}{\Vec{r}}

\newcommand{\pPderivT}{\partial_{\pTime}}   
\newcommand{\pVderiv}{\Vec{\nabla}}      

\newcommand{\pPermittivityVac}{\epsilon_0} 

\newcommand{\pXemXelectric}{E}                      
\newcommand{\pXemXelectricv}{\Vec{\pXemXelectric}}  
\newcommand{\pXemXdisplacement}{D}                         
\newcommand{\pXemXdisplacementv}{\Vec{\pXemXdisplacement}} 
\newcommand{\pXemXpolarization}{P}                         
\newcommand{\pXemXpolarizationv}{\Vec{\pXemXpolarization}} 

\newcommand{\pXemXspotential}{\phi}        

\newcommand{\pCurrent}{J}                  
\newcommand{\pCurrentv}{\Vec{\pCurrent}}   

\newcommand{\pEfield}{\pXemXelectric}

\newcommand{\pEfieldv}{\pXemXelectricv}
\newcommand{\pDfieldv}{\pXemXdisplacementv}
\newcommand{\pPfieldv}{\pXemXpolarizationv}

\def\Pfield{\pXemXpolarizationv}
\def\wPlasma{\omega_{\textup{P}}}

\def\Qfield{{\Psi}}   
\def\qfield{{\psi}}   
\def\ppphi{\pXemXspotential} 

\def\plloss{\gamma}          
\def\plspeed{\beta}          
\def\pldriving{\alpha}       
\def\plratio{\sigma}         

\def\Jfield{\pCurrentv}
\def\Nfield{N}               
\def\nfield{\eta}            

\begin{document}
\title{The deconstructed hydrodynamic model for plasmonics}

\author{Paul Kinsler}
\homepage[]{https://orcid.org/0000-0001-5744-8146}
\email[\hphantom{.}~]{Dr.Paul.Kinsler@physics.org}
\affiliation{
  Physics Department,
  Lancaster University,
  Lancaster LA1 4YB, 
  United Kingdom,
}
\affiliation{
  Cockcroft Institute, 
  Sci-Tech Daresbury, 
  Daresbury WA4 4AD, 
  United Kingdom.
}

\lhead{\includegraphics[height=5mm,angle=0]{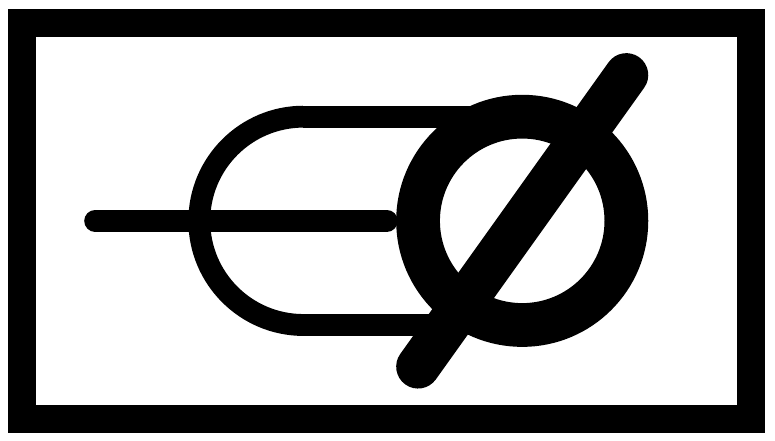}~~DEHYDRO}
\chead{Deconstructed Hydrodynamic Plasmonics}
\rhead{
\href{mailto:Dr.Paul.Kinsler@physics.org}{Dr.Paul.Kinsler@physics.org}\\
\href{http://www.kinsler.org/physics/}{http://www.kinsler.org/physics/}
}

\begin{abstract}

I give an exact but deconstructed version of the 
 second-order wave-like equation that encapsulates the 
 hydrodynamic model for plasmonics.
Comprising two first order equations, 
 the deconstruction has potential uses in understanding or interpreting 
 the hydrodynamic model, 
 since its meaning is not obscured by approximation.
However, 
 as the physical interpretation of the deconstructed model
 is difficult, 
 due to the choice of the polarization as the significant quantity, 
 I also consider a alternate model 
 based on the polarization current.
This alternate model has a clear and direct physical interpretation.

\end{abstract}


\date{\today}
\maketitle
\thispagestyle{fancy}

%

%
\section{Introduction}\label{S-intro}

The hydrodynamic model for plasmonics (HMP)
 \cite{Ciraci-PS-2013cphc}
 has recently found widespread popularity in the field of plasmonics.
One of its key features is that
 unlike simpler plasmonics approaches based on the 
 Drude model, 
 it incorporates spatial derivatives terms 
 which represent the dynamics of the charge distribution.
Although these are often called ``non-local'' effects, 
 they are more usefully called propagation effects, 
 since they are not non-local in any sense that 
 violates relativistic (signalling) constraints on the physics.
However, 
 although much of its usage in plasmonics is recent, 
 the basic model itself dates back to the 1970's \cite{Eguiluz-Q-1976prb}
 and has been used in a number of other contexts 
 \cite{Fedorov-BRG-2005prb,Baranov-FRM-2003prb}.

Since it is most efficient to direct the reader interested
 in the physical basis and assumptions of the HMP
 to the motivation,
 derivation,
 and exposition of Ciraci et. al \cite{Ciraci-PS-2013cphc}; 
 here we will simply repeat their relevant equation (14),
 which describes the how the wave-like response of the charge distribution
 in the material appears as a standard electromagnetic polarization.
It is
~
\begin{align}
  \pPderivT^2 \Pfield
 +
  \plloss \pPderivT \Pfield
 -
  \pVderiv \plspeed^2 \pVderiv \cdot \Pfield
&=
  \pPermittivityVac
  \wPlasma^2
  \pEfieldv
,
\label{eqn-CPShydrodynamicmodel}
\end{align}
where we should remember that 
 the polarization field $\Pfield \equiv \Pfield(\pTime,\pSpace)$
 and the driving electric field $\pEfieldv \equiv \pEfieldv (\pTime,\pSpace)$.
In the following I prefix all equation numbers 
 from this reference with CPS, 
 so that the above is then (CPS14).
In this equation we note that the parameter $\plspeed$
 is defined by 
~
\begin{align}
  \plspeed^{2}
=
  \frac{2E_F}{3m_e}
\end{align}
 where $E_F$ is the Fermi energy
 and
 $m_e$ is the electron mass.
Also, 
 $\plloss$ is the polarization decay rate,
 $\pPermittivityVac$ is the permittivity of vacuum,
 $\wPlasma = \sqrt{n_0 e^2 / \pPermittivityVac m_e}$ 
 is the plasma frequency, 
 $n_0$ is the background electron number density,
 and $e$ the electron charge.
Key features of the derivation are
 (a) the substitution of
 the polarization $\Pfield$ for the (polarization) current $\Jfield$,
 (b) the dropping of all but the dominant term in the Lorentz force equation, 
 and 
 (c) assuming that the fluctuations in the polarization and/or 
  charge densitys are small.
Here step (a) in particular results in the above equation for $\Pfield$, 
 a useful feature when we plan to link the plasmonic behaviour 
 to that of the electric field, 
 since it allows us to define an effective permittivity.
However, 
 the replacement of $\Jfield$ is not strictly necessary,
 and the derivation can proceed without it.

For the purposes of my discussion here,
 it is important to note that 
 this model is not exactly equivalent to the starting assumptions 
 used in its derivation.
While this is of course completely natural, 
 given the approximations made in the derivation, 
 and is indeed in itself an unremarkable statement, 
 given the ubiquity of the role of approximations in physical model-making, 
 we might still ask the question:
 What is this approximate HMP model \emph{exactly} equivalent to?

In Section \ref{S-Deconstruction} I present a physical model which \emph{is} 
 exactly equivalent to the HMP, 
 and discuss some of the implications, 
 while in Section \ref{S-Jdeconstruction} I show how an 
 alternative form based on the polarization current $\Jfield$
 has a simpler physical interpretation.
Lastly, 
 in Section \ref{S-conclusion} I summarize the results.

%
\section{Deconstruction of the HMP}\label{S-Deconstruction}

We can deconstruct the HMP into two first order pieces, 
 one defining how (scaled) charge density gradients 
 drive changes in the dielectric polarization, 
 along with losses; 
 and the other defining how the divergence of the polarization
 drives changes in a quantity related to the charge density, 
 along with the effect of the electric field potential.

The deconstructed HMP (D-HMP) equations are
~
\begin{align}
  \pPderivT \Pfield
&=
  \pVderiv \qfield
 -
  \plloss \Pfield
\label{eqn-dtP-firstorder}
\\
  \pPderivT \Qfield
&=
  \pVderiv \cdot \Pfield
 +
  \pldriving \ppphi
\label{eqn-dtQ-firstorder}
\end{align}
 where $\Qfield(\pTime,\pSpace) = \plratio \qfield(\pTime,\pSpace)$
 and there is a scalar electric potential 
 $\pEfieldv(\pTime,\pSpace) = \pVderiv\ppphi(\pTime,\pSpace)$.
Although we could have put the driving electric field in 
 \eqref{eqn-dtP-firstorder}, 
 it would have to have been in the form of the time-integral of the field, 
 and so disrupting the causal interpretation of the equation
 \cite{Kinsler-2011ejp}.

Note that these two equations are very similar to 
 a modified and reinterpreted version 
 of ``p-acoustics'' 
 \cite{Kinsler-M-2014pra,Kinsler-M-2015raytail,McCall-K-2017hbookmm}, 
 albeit with extra driving terms.

\subsection{Equivalence}\label{S-equivalence}

To demonstrate that these two coupled first-order D-HMP equations
 \emph{are} in fact equivalent to the second-order HMP equation, 
 we can combine them.
We first take the time derivative of \eqref{eqn-dtP-firstorder}, 
 then substitute in \eqref{eqn-dtQ-firstorder}, 
 so that 
~
\begin{align}
  \pPderivT^2 \Pfield
&=
  \pVderiv \pPderivT \qfield
 -
  \plloss \pPderivT \Pfield
\\
&=
  \pVderiv \pPderivT \plratio^{-1} \Qfield
  -
  \plloss \pPderivT \Pfield
\\
&=
  \pVderiv \plratio^{-1} \pPderivT \Qfield
  -
  \plloss \pPderivT \Pfield
\\
&=
  \pVderiv \plratio^{-1} \pVderiv \cdot \Pfield
 +
  \pVderiv \plratio^{-1} \pldriving \ppphi
  -
  \plloss \pPderivT \Pfield
\\
&=
  \pVderiv \plratio^{-1} \pVderiv \cdot \Pfield
 +
  \plratio^{-1} \pldriving \pVderiv \ppphi
  -
  \plloss \pPderivT \Pfield
\\
&=
  \pVderiv \plratio^{-1} \pVderiv \cdot \Pfield
 +
  \plratio^{-1} \pldriving \pEfieldv
  -
  \plloss \pPderivT \Pfield
.
\label{eqn-d2tP-DHMP}
\end{align}
In doing this we have necessarily made some assumptions, 
 namely that $\plratio$ is a fixed paramenter 
 with no time or space dependence, 
 and that $\pldriving$ has no space dependence.
These assumptions are unremarkable, 
 since both the original HMP model and this one 
 are predicated on being in a homogeneous background.
However, 
 with this new D-HMP, 
 we \emph{could} allow them to vary and so derive a more general version 
 of \eqref{eqn-d2tP-DHMP}.

Although \eqref{eqn-d2tP-DHMP}
 is structurally similar to \eqref{eqn-CPShydrodynamicmodel}, 
 to show they are identical we need to fix the parameters
 $\plratio$ and $\pldriving$.
By comparing terms, 
 we straightforwardly find that 
~
\begin{align}
  \plratio
&=
  \plspeed^{-2}
=
  \frac{3m_e}{2E_F}
,
\end{align}
 and 
~
\begin{align}
  \plratio^{-1} \pldriving
&=
  \pPermittivityVac \wPlasma^2
=
  \frac{3 n_0 e^2}{2E_F}
.
\end{align}

Hence our D-HMP equations have combine to form exactly the 
 HMP model equation from \eqref{eqn-CPShydrodynamicmodel}, 
 i.e. 
~
\begin{align}
  \pPderivT^2 \Pfield
 +
  \plloss \pPderivT \Pfield
 -
  \pVderiv \plspeed^2 \pVderiv \cdot \Pfield
&=
  \pPermittivityVac
  \wPlasma^2
  \pEfieldv
.
\label{eqn-PKhydrodynamicmodel}
\end{align}

Now, 
 let us briefly reiterate the distinction between the 
 two models, 
 a difference which exists despite the fact that 
 \eqref{eqn-CPShydrodynamicmodel} and \eqref{eqn-PKhydrodynamicmodel}
 are identical:

\begin{itemize}

\item 
 Eqn. \eqref{eqn-CPShydrodynamicmodel}
 is an \emph{approximation} to the HMP's starting point, 
 which considers the dynamics of an electron fluid dynamics
 on an otherwise homogeneous background,
 i.e. (CPS3) and following equations.

\item
 Eqn. \eqref{eqn-PKhydrodynamicmodel}
 is an exact counterpart to the D-HMP's starting point
 of coupled  polarization and charge density equations
 on an otherwise homogeneous background,
 i.e. \eqref{eqn-dtP-firstorder},
 \eqref{eqn-dtQ-firstorder}.

\end{itemize}

%
\subsection{Interpretation}\label{S-Interpretation}

The two equations \eqref{eqn-dtP-firstorder},
 \eqref{eqn-dtQ-firstorder}
 are not seen in the usual derivation \cite{Ciraci-PS-2013cphc}.
The distinction is primarily due to the fact that 
 the HMP starting point in \cite{Ciraci-PS-2013cphc}
 focusses on charge density, 
 charge fluid velocity, 
 and current; 
 whereas the result looks at polarization $\Pfield$, 
 which is the \emph{time integral} of the current.
This time-integral property of the polarization
 can make interpretations somewhat tricky\footnote{See also
   e.g. Faraday's Law \cite{Kinsler-2017arxiv-faradin}}, 
 as here, 
 although the drawbacks are usually minor
 compared to the advantage it provides 
 in allowing us to use it as an input to macroscopic electrodynamics, 
 i.e. in writing $\pDfieldv = \pPermittivityVac \pEfieldv + \pPfieldv$.

Since the D-HMP is a deconstruction of the HMP result, 
 this means that the other D-HMP quantities are, 
 like polarization, 
 also time integrated quantities --
 hence the appearance of the scalar electric potential $\ppphi$
 in place of the electric field $\pEfieldv$.
The $\Qfield, \qfield$ quantities are likewise related
 to the time integral of the number and/or charge densities, 
 which might therefore be thought of as (called) ``accumulations''
 rather than densities.

The first equation \eqref{eqn-dtP-firstorder}
 is force-law like:
~
\begin{align}
  \pPderivT \Pfield
&=
  \pVderiv \qfield
 -
  \plloss \Pfield
,
\nonumber
\end{align}
 with changes in polarization $\Pfield$
 following from gradients in the 
 potential-like accumulation field $\qfield$;
 and from linear losses proportional to $\plloss$.

The second equation \eqref{eqn-dtQ-firstorder}
 is conservation-law like:
~
\begin{align}
  \pPderivT \Qfield
&=
  \pVderiv \cdot \Pfield
 +
  \pldriving \ppphi
,
\nonumber
\end{align}
 with changes in the the accumulation field $\Qfield$
 following from the divergence --
 local inflows or outflows --
 of the polarization; 
 whilst also being augmented by driving from the electric scalar potential.

The fact that this conservation equation is for this 
 new (and perhaps somewhat mysterious)
 accumulation field $\Qfield$,  
 and further, 
 is driven by the electric potential rather than the electric field,
 is a result of the requirement for a dynamic equation for $\pPfieldv$
 rather than a well-defined microscopic quantity 
 such as (e.g.) current $\pCurrentv$.

%
\section{Current-based form: the J-HMP}\label{S-Jdeconstruction}

We have seen above that the interpretation of the standard D-HMP model
 is physically rather unsatisfactory.
However, 
 this situation can be avoided
 by instead using an equation for
 the polarization current $\Jfield(\pTime,\pSpace)$,
 rather than its time integral, 
 the polarization $\Pfield(\pTime,\pSpace)$ itself.
We can generate such an expression by simply
 taking the time-derivative of \eqref{eqn-CPShydrodynamicmodel}
 and substituting $\Jfield$ for $\pPderivT\Pfield$;
 but for the interested reader it is nevertheless 
 worthwhile to revisit the derivation of 
 Ciraci et. al \cite{Ciraci-PS-2013cphc}, 
 and seeing that it still follows without
 the substitution. 
The modified current-based version of 
 (CPS14) and \eqref{eqn-CPShydrodynamicmodel}
 is 
~
\begin{align}
  \pPderivT^2 \Jfield
 +
  \plloss \pPderivT \Jfield
 -
  \pVderiv \plspeed^2 \pVderiv \cdot \Jfield
&=
  \pPermittivityVac
  \wPlasma^2
  \pPderivT \pEfieldv
.
\label{eqn-J-CPShydrodynamicmodel}
\end{align}

We can deconstruct this ``J-HMP'' into two first order pieces, 
 one defining how charge density gradients 
 drive changes in the polarization current, 
 along with losses; 
 and the other defining how the divergence of the polarization current
 drives changes in the charge density.

The deconstructed J-HMP (DJ-HMP) equations are
~
\begin{align}
  \pPderivT \Jfield
&=
  \pVderiv \nfield
 -
  \plloss \Jfield
 +
  \pldriving'
  \pEfieldv
\label{eqn-J-dtP-firstorder}
\\
  \pPderivT \Nfield
&=
  \pVderiv \cdot \Jfield
\label{eqn-J-dtQ-firstorder}
\end{align}
 where $\Nfield(\pTime,\pSpace) = \plratio \nfield(\pTime,\pSpace)$.
Here $\Nfield$ is a linear polarization charge density 
 (in units $C/m$),
 and $\nfield$ is a closely related ``current velocity'' quantity
 (in units $C.m/s^2$).

Note the different handling of the driving field in this version, 
 which can now act straightforwardly on the polarization current
 without introducing problems.

\subsection{Equivalence}\label{S-Jequivalence}

To demonstrate that these two coupled first-order DJ-HMP equations
 \emph{are} in fact equivalent to the second-order J-HMP equation, 
 we can combine them.
We first take the time derivative of \eqref{eqn-dtP-firstorder}, 
 then substitute in \eqref{eqn-dtQ-firstorder}, 
 so that 
~
\begin{align}
  \pPderivT^2 \Jfield
&=
  \pVderiv \pPderivT \nfield
 -
  \plloss \pPderivT \Jfield
 +
  \pldriving'
  \pPderivT \pEfieldv
\\
&=
  \pVderiv \pPderivT \plratio^{-1} \Nfield
  -
  \plloss \pPderivT \Jfield
 +
  \pldriving'
  \pPderivT \pEfieldv
\\
&=
  \pVderiv \plratio^{-1} \pPderivT \Nfield
  -
  \plloss \pPderivT \Jfield
 +
  \pldriving'
  \pPderivT \pEfieldv
\\
&=
  \pVderiv \plratio^{-1} \pVderiv \cdot \Jfield
  -
  \plloss \pPderivT \Jfield
 +
  \pldriving'
  \pPderivT \pEfieldv
.
\label{eqn-J-d2tP-DHMP}
\end{align}
Here we have, 
 as previously, 
 assumed that $\plratio$ is a fixed paramenter 
 with no time or space dependence, 
 and also that $\pldriving'$ is independent of time.
These assumptions are unremarkable, 
 since both the original HMP model, 
 the J-HMP one, 
 and this one 
 are predicated on being in a homogeneous background.
However, 
 with this new DJ-HMP, 
 just as with the D-HMP,
 we \emph{could} allow them to vary and so derive a more general version 
 of \eqref{eqn-J-d2tP-DHMP}.

Although \eqref{eqn-J-d2tP-DHMP}
 is structurally similar to \eqref{eqn-J-CPShydrodynamicmodel}, 
 to show they are identical we need to fix the parameters
 $\plratio$ and $\pldriving'$.
Fortunately these are the same as for the D-HMP model, 
 but with a driving strength $\pldriving'=\plratio^{-1}\pldriving$.
Hence our DJ-HMP equations combine to form exactly the 
 J-HMP model equation from \eqref{eqn-J-CPShydrodynamicmodel}, 
 i.e. 
~
\begin{align}
  \pPderivT^2 \Jfield
 +
  \plloss \pPderivT \Jfield
 -
  \pVderiv \plspeed^2 \pVderiv \cdot \Jfield
&=
  \pPermittivityVac
  \wPlasma^2
  \pEfieldv
.
\label{eqn-J-PKhydrodynamicmodel}
\end{align}

The distinctions between the J-HMP model 
 and the DJ-HMP model are exactly analogous to the 
 distinctions between the HMP model and the D-HMP one.

%
\subsection{Interpretation}\label{S-Jinterpretation}

Examining the DJ-HMP model equations
 \eqref{eqn-J-dtP-firstorder} 
 and
 \eqref{eqn-J-dtQ-firstorder}
 we see that we now have a very much more straightforward interpretation.

The first equation \eqref{eqn-J-dtP-firstorder}
 is force-law like:
~
\begin{align}
  \pPderivT \Jfield
&=
  \pVderiv \nfield
 -
  \plloss \Jfield
 +
  \pldriving'
  \pEfieldv
,
\nonumber
\end{align}
 with changes in polarization current $\Jfield$
 following from gradients in the scaled charge density
 (i.e. $\nfield = \plratio^{-1} \Nfield$),
 from linear losses proportional to $\plloss$,
 and the driving effect of the electric field $\pEfieldv$

The second equation \eqref{eqn-J-dtQ-firstorder}
 is conservation-law like:
~
\begin{align}
  \pPderivT \Nfield
&=
  \pVderiv \cdot \Jfield
,
\nonumber
\end{align}
 with changes in the the polarization charge density $\Nfield$
 following from the divergence --
 local inflows or outflows --
 of the polarization current.

The single drawback of this form is that, 
 without a direct expression for $\Pfield$, 
 it is not as simple to construct the effective permittivity
 resulting from the plasmonic response.
However, 
 it is easy enough -- we just have to time-integrate
 the polarization current $\Jfield$; 
 either that,
 or take a more microscopic view of solving Maxwell's equations.

%
\section{Conclusion}\label{S-conclusion}

The proposal made here is that 
 when trying to understand the behaviour of the HMP, 
 it can be useful to (first) understand it in terms 
 of the D-HMP.
This is because the D-HMP equations do not suffer 
 from the conflated effects of the serial approximations 
 used in the HMP derivation.
While a discussion of the behavior of the HMP model might easily
 get bogged down over ambiguities introduced by
 any of the approximations utilised in the HMP derivation, 
 this possibility is avoided in the D-HMP equations.

However, 
 we also see that the D-HMP equations have
 a somewhat obscure physical interpretation.
This lead us to reformulate the plasmonic response in terms of 
 the polarization current
 in the DJ-HMP model,
 an uncomplicated procedure that clarified the physical meaning, 
 but left us without a direct expression for the behaviour 
 of the polarization.

%


%
\section*{Appendix: HMP for $\Psi$}\label{S-appendix}


As an exercise, 
 I do the substitution in the reverse order to 
 derive a second order wave-like equation for $\Qfield$.
We have
~
\begin{align}
  \partial_t^2 \Qfield
&=
  \pVderiv \cdot \partial_t \Pfield
 +
  \pldriving \partial_t \ppphi
\\
&=
  \pVderiv \cdot \left( \pVderiv \qfield - \plloss \Pfield \right)
 +
  \pldriving \partial_t \ppphi
\\
&=
  \plratio^{-1} \pVderiv \cdot \pVderiv \Qfield
 -
  \plloss \pVderiv \cdot \Pfield 
 +
  \pldriving \partial_t \ppphi
\\
&=
  \plratio^{-1} \pVderiv \cdot \pVderiv \Qfield
 -
  \plloss \left( \pPderivT \Qfield - \pldriving \ppphi \right)
 +
  \pldriving \partial_t \ppphi
\\
&=
  \pVderiv \cdot \plspeed^2 \pVderiv \Qfield
 -
  \plloss \pPderivT \Qfield
 +
  n_0 m_e
  \left(\frac{3e}{2E_F}\right)^2
  \left(
    \partial_t \ppphi
   +
    \plloss \ppphi
  \right)
.
\end{align}
Not unexpectedly, 
 we see the same phase velocity $\plspeed$.
Less conveniently, 
 the driving term is no longer simply dependent on $\pEfieldv$, 
 depending instead on a combination of its time derivative 
 and a loss-dependent fraction of itself.

However, 
 if the driving electric field is CW
 at a frequency $\omega_0$ and wavevector $\Vec{k}$,
 so that 
~
\begin{align}
  \ppphi 
&=
  \ppphi_0 
  \exp 
  \left[
    \imath 
    \left( \omega_0 t - \Vec{k} \cdot \Vec{r} \right)
  \right]
.
\end{align}

Along the propagation direction,
 we know that $k$ is the spatial derivative of the potential, 
 so that with $\pEfield_0 = k \ppphi_0$
~
\begin{align}
  \pEfield
&=
  \pEfield_0 
  \exp 
  \left[
    \imath 
    \left( \omega_0 t - \Vec{k} \cdot \Vec{r} \right)
  \right]
.
\end{align}

Thus simplifying and assuming that propagation is along $x$
 with $\bar{\pldriving} ={\pldriving} \left|\imath \omega_0 + \plloss\right| /  k$,
 and absorbing any phase into $\varphi$,
~
\begin{align}
  \partial_t^2 \Qfield
&=
  \pVderiv \cdot \plspeed^2 \pVderiv \Qfield
 -
  \plloss \pPderivT \Qfield
 +
  \imath 
  \pldriving 
  \left(
    \imath \omega_0
   +
    \plloss
  \right)
  k^{-1} \pEfield_0 e^{\imath \left(\omega_0 t - k x + \varphi\right)}
\\
&=
  \pVderiv \cdot \plspeed^2 \pVderiv \Qfield
 -
  \plloss \pPderivT \Qfield
 +
  \bar{\pldriving}
  \pEfield_0 e^{\imath \left(\omega_0 t - k x + \varphi\right)}
.
\end{align}

\end{document}